\documentstyle[twocolumn,epsf,prl,aps]{revtex}
\newcommand {\be}{\begin{equation}}
\newcommand {\ee}{\end{equation}}
\newcommand {\bea}{\begin{eqnarray}}
\newcommand {\eea}{\end{eqnarray}}
\usepackage{graphicx}
\begin{document}

\draft

\title
{Magnetoconductivity due to Quantum 
Interference in Strongly Underdoped  $YBa_2Cu_3O_{x}$  } 
\author{  E. Cimpoiasu, G. A. Levin, and C. C. Almasan}
\address{Department of Physics, Kent State University, Kent  OH 44242}
\author{  Hong Zheng and B. W. Veal}
\address{Materials Science Division, Argonne National Laboratory, Argonne,
IL  60439
\vspace{0.5cm}
{\rm \begin{quote}
We report magnetoconductivity measurements on $YBa_2Cu_3O_{x}$ ($x=6.25$ and $6.36$) single
crystals. Our main result  is  that both the in-plane $\Delta\sigma_{ab}$
and out-of-plane  $\Delta\sigma_{c}$ magnetoconductivities exhibit the  field dependence characteristic of
"two-dimensional" quantum interference  in applied magnetic fields $H ||c$. 
Namely,  $ \Delta\sigma_{c,ab}\propto \ln H/H_0 >0 $, with
$\Delta\sigma_{c}/\sigma_{c}$ substantially greater than
$\Delta\sigma_{ab}/\sigma_{ab}$. We interpret this result as an evidence of
interlayer incoherence in these crystals, so that the phase-coherent trajectories are mostly confined to one
bilayer. 
\end{quote}
\date{\today}
}
}
\maketitle

\section{Introduction}

Understanding  the anomalous features  of the in-plane  and out-of-plane
normal state transport  in layered cuprates remains a challenge. 
Among the properties of the resistivity tensor $\{\rho_{c},\rho_{ab}\}$ that are some of the most unusual from the
point of view of the conventional Fermi liquid theory are strongly temperature dependent anisotropy 
$\rho_c/\rho_{ab}$ and coexistence of metallic $\rho_{ab}(T)$ and nonmetallic $\rho_{c}(T)$ over extended
temperature and doping ranges.  It has been suggested that these anomalies of the out-of-plane transport 
are the result of interlayer incoherence\cite{Anderson,Graf}. 
This means that the phase coherent trajectories along which the single electrons maintain their phase
memory are all confined to a single bilayer because the interbilayer transitions lead to dephasing 
of still unknown origin.  An experimental confirmation of c-axis incoherence was 
provided by optical measurements\cite{Basov}. 

The suggestion of interlayer incoherence in the normal state of the  cuprates, however, is
not fully accepted. There are  alternative models of
the out-of-plane transport that do not involve interlayer incoherence\cite{Pines,Abrikosov1}. 
The phase coherent trajectories  in such models are three dimensional and extend over many unit cells in
both the in- and out-of-plane directions. 

Magnetoeffects due to quantum interference allow, in principle, the determination of the dimensionality of the phase
coherence. Magnetoconductivity arises in relatively small magnetic fields due to contributions of
the self-intersecting phase coherent trajectories along which the loops can be traversed in two different
directions\cite{Altshuler,Lee,Abrikosov}. The magnitude and field dependence of magnetoresistivity 
are determined 
by the probability of a trajectory to form a loop of a given area. 
If the phase-coherent trajectories are  two-dimensional 2D, they are  substantially
more likely to form large loops than three-dimensional 3D ones. This results in a more pronounced effect,
observable at higher temperatures. The field dependence of magnetoconductivity is also different in these two
cases; namely, 2D trajectories lead to 
$\Delta\sigma \propto \ln H$,
while 3D trajectories lead to $\Delta\sigma \propto H^{1/2}$.

Here we present magnetoresistivities of two strongly underdoped 
$YBa_2Cu_3O_{6.25}$ and $YBa_2Cu_3O_{6.36}$ single crystals.  Our results can be summarized as follows:
Both components 
$\Delta\rho_c/\rho_c$ and $\Delta\rho_{ab}/\rho_{ab}$ of the magnetoresistivity tensor reveal  the presence of two different
conduction mechanisms in magnetic fields $H\|c$. (1) In low fields,  the  
magnetoresistivities are negative and have a logarithmic field dependence 
over about a decade (at the lowest accessible temperature). This behavior represent the main focus of this
paper. What distinguishes these results from traditional experiments on ultrathin films and 
other 2D systems \cite{Altshuler,Kawaguchi}, is
that the out-of-plane  magnetoresistivity has the logarithmic field dependence and is substantially larger
than the in-plane  magnetoresistivity. We interpret these findings as a confirmation of a strong interbilayer
decoherence and, therefore,  the two-dimensional nature of the phase-coherent paths. Under this condition, both
components of the conductivity depend on the in-plane phase coherence length $\ell_{\varphi }$,
which changes with
field logarithmically, thus  producing the corresponding $H$-dependence of the magnetoresistivities. (2) In high
magnetic fields and at low temperatures, both magnetoresistivities become positive and change as 
$\gamma H^2$ up to the largest available field of $14\;T$. This contribution arises from
antiferromagnetic correlations. 

\section{Experimental Details}
Strongly underdoped
$YBa_2Cu_3O_{x}$ ($x=6.36$ and
$6.25$) single crystals were grown in gold crucibles using the self-flux
method. The oxygen stoichiometry was adjusted by annealing the samples at $500^{o} C $ in a predetermined
$O_2-N_2$ atmosphere \cite{Veal}.  In-plane $\rho_{ab}$ and out-of-plane
$\rho_c$ resistivities and the respective magnetoresistivities (MR)  were measured by a 
multiterminal
method on the same single crystal,  in magnetic fields
$H$ up to
$14\;T$. This allows us to carry out a {\it quantitative} comparison  between 
in-plane $\Delta\rho_{ab}/\rho_{ab}$  and out-of-plane $\Delta\rho_c/\rho_c$  magnetoresistivities as a function of
temperature
$T$ and applied magnetic field $H$.  All previous measurements of MR in $YBa_2Cu_3O_{x}$ reported in the
literature were obtained by a four-point method, so that  $\Delta\rho_{ab}/\rho_{ab}$ and
$\Delta\rho_c/\rho_c$ were measured on different single crystals. Since magnetoresistivity is very
small, inevitable variations of stoichiometry, shape, and size of specimens  have a large effect on its
value and preclude any quantitative analysis of correlations between different components of the
magnetoresistivity tensor.

Low contact resistance gold leads were attached using thermally treated silver pads 
and room temperature silver epoxy. We applied an ac  electrical current  
$I\sim 0.1\; mA $ through the  two leads on the "top" face and  measured the two
voltage drops on the same  (top voltage $V_t$) and the opposite  (bottom voltage $V_b$) faces 
of the sample \cite{Jiang}.
The two voltages $V_{t,b}$ and their variation with field $\delta V_{t,b}$ were measured using a
low-frequency 17 Hz resistance bridge  at constant  temperature, while  sweeping  the magnetic field H applied
parallel to the c-axis of the single crystal. 

The temperature  of the sample chamber was kept constant using two
temperature sensors ($Pt $ thermometer for $T > 100\ K$ and $Cx $ sensor for $T < 100\ K$). The 
temperature change  $\delta T(H)$  resulting from the magnetoresistance of the sensors
was  measured by carefully calibrating them using a capacitance 
thermometer, which has zero magnetoresistance. 
The effect of the magnetic field on $ V_{t,b} $  was then
determined by subtracting the effect of this temperature change from the raw 
data; i.e.,
\be
\Delta V_{t,b} (H) = \delta V_{t,b} - \delta T(H)\frac{dV_{t,b}}{dT}.
\ee

Unlike in the four-point method, the distribution of the current inside 
the sample is nonuniform. As a result, the Hall voltage
contributes to the field variation of the top and bottom voltages. To first order in
magnetoconductivities, appropriate for these measurements,
\be
\Delta V_{t,b} (H) = A_{t,b}\Delta\sigma_{xx} +B_{t,b}\Delta\sigma_{zz} +C_{t,b}\sigma_{xy},
\ee
where $\Delta\sigma_{xx}$ and $\Delta\sigma_{zz}$ are magnetoconductivities, while
$\sigma_{xy}$ is the off-diagonal component of the conductivity tensor. The coefficients 
$A_{t,b},\;B_{t,b}$ and $C_{t,b}$ can be obtained from the solution of Laplace's equation 
determining the distribution of voltage inside the sample. (Within a linear approximation,
these coefficients depend on the  zero-field values of the conductivity tensor, which is diagonal.)

The most reliable way to determine the magnetoconductivities  $\Delta\sigma_{xx}$ and
$\Delta\sigma_{zz}$ from Eq. (2) is to eliminate the contribution of the  Hall effect by measuring
$\Delta V_{t,b} (H)$ at two opposite polarities
$\pm H$ of the applied magnetic field.  The even component
$\Delta V_{t,b}^{+} (H)= 1/2[\Delta V_{t,b}(H)+\Delta V_{t,b}(-H)]$ determines the magnetoconductivity,
while the odd component $\Delta V_{t,b}^{-} (H)= 1/2[\Delta V_{t,b}(H)-\Delta
V_{t,b}(-H)]=C_{t,b}\sigma_{xy}$. The magnetoconductivities were then calculated from $\Delta V_{t,b}^{+}
(H)$ using  an   algorithm  described in\cite{Jiang}. 

As an example of the raw data, the main panel of Fig. 1 shows the field dependence of $\Delta V_{t}$ of the $x=
6.36$ single crystal, measured at $ T = 100\;K$, while the inset displays its odd
$\Delta V_{t}^{-}$ and even $\Delta V_{t}^{+}$ components. The  field dependence of $\Delta V_{t}$ clearly
separates into quadratic dependence of $\Delta V_{t}^{+}$ and, consequently, of  magnetoconductivity,
and linear dependence of $\Delta V_{t}^{-}$ and, hence, Hall conductivity $\sigma_{xy}$.
\begin{figure}
\epsfxsize=\columnwidth
\epsfbox{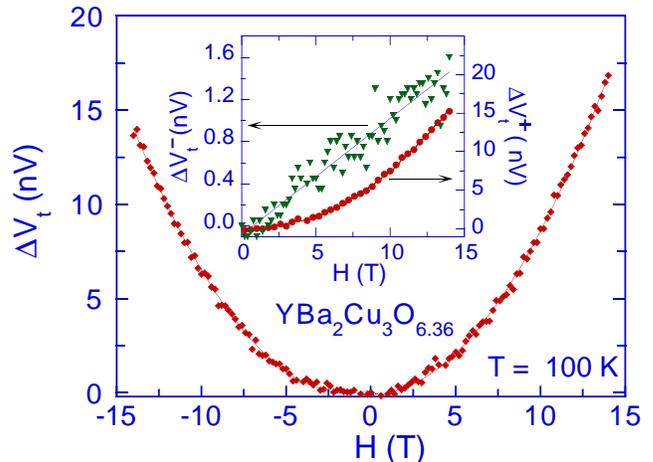}
\caption{ Main panel: The field $H$ dependence of the top voltage $\Delta V_{t}$ for the $x=6.36$ single crystal,
 measured at $ T = 100\;K$ in $H\parallel c$; Inset:  odd
$\Delta V_{t}^{-}=1/2[\Delta V_{t}(H)-\Delta V_{t}(-H)]$ and even $\Delta V_{t}^{+}=1/2[\Delta V_{t}(H)+\Delta V_{t}(-H)]$
components of the top  voltage obtained from the data shown in the main panel.}
\label{}
\end{figure}

The multiterminal technique has never been used before to measure magnetoconductivities of cuprates. 
The application of this method  to magneto-measurements has subtle, but important, differences with the
conventional four-point method. For example, in the multiterminal method
one measures magnetoconductivities, while in the conventional four-point method one measures 
magnetoresistivities. It should be noted that most theoretical approaches to magnetoeffects describe
magnetoconductivities rather than magnetoresistivities. Assuming that 
$\sigma_{xz}$ and $\sigma_{yz}$ are negligible, the relationship between
magnetoresistivities and magnetoconductivities is given by\cite{Pippard}:
\be
\frac{\Delta\rho_{xx}}{\rho_{xx}}=-\frac{\Delta\sigma_{xx}}{\sigma_{xx}}-
\frac{\sigma_{xy}^2}{\sigma_{xx}^2};\;\;
\frac{\Delta\rho_{zz}}{\rho_{zz}}=-\frac{\Delta\sigma_{zz}}{\sigma_{zz}}
\ee
Comparing the values of the even  and odd components $\Delta V_{t,b}^{\pm}$, 
we  find that in most cases $\Delta\sigma_{xx}\sim \sigma_{xy}$ and, since both of them are very small,
the quadratic term in Eq. (3) is negligible.  Therefore, the magnetoresistivities are given directly by
magnetoconductivities with inverted sign.   The only exception to this case is at very low magnetic fields where
$\sigma_{xy}>\Delta\sigma_{xx}$ since $\sigma_{xy}\propto H$ while
$\Delta\sigma_{xx}\propto H^2$. Therefore, at very low fields, one needs to know $\sigma_{xy}$ 
in order to obtain magnetoresistivity from magnetoconductivity. 
In this paper, we present all our results in terms of magnetoresistivity obtained from the measured
magnetoconductivity according to Eq. (3) where we neglect the quadratic term. This facilitates a direct
comparison between our data and already published magnetoresistivity data.

\section{Magnetoresistivity }
Figures 2(a) and 2(b) show the T-dependences of the resistivities of $YBa_2Cu_3O_{6.36}$ and
$YBa_2Cu_3O_{6.25}$ single crystals, respectively, in zero magnetic field. These single crystals display
a diverse $T-$dependence of resistivities, and antiferromagnetism below
$T\approx40\;K$ in $x=6.36$ sample and for all $T< 300\;K$ in $x=6.25$ crystal.
Coexistence of metallic
$\rho_{ab}$  and nonmetallic $\rho_{c}$ over an extensive temperature range is characteristic of
underdoped cuprates. We note that the in-plane resistivity  $\rho_{ab}$ is 
metallic  above a certain temperature even in the strongly underdoped samples.  The crossover  to  
nonmetallic behavior in $\rho_{ab}$ takes place at progressively higher temperatures 
in  more underdoped samples. For 
\begin{figure}
\epsfxsize=\columnwidth
\epsfbox{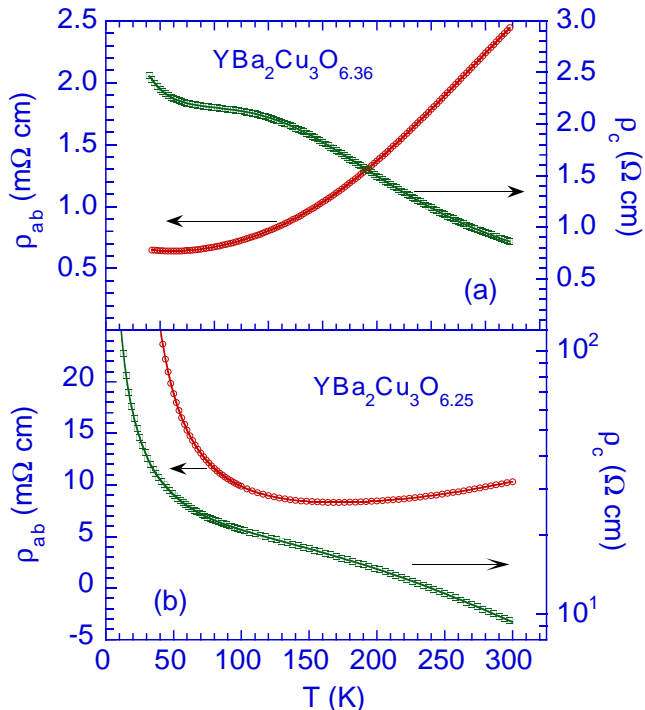}
\caption{Temperature $T$ dependence  of  zero-field  in-plane $\rho_{ab}$ and 
out-of-plane $\rho_{c}$ resistivities for (a) $YBa_{2}Cu_{3}O_{6.36}$, and (b)
$YBa_{2}Cu_{3}O_{6.25}$.}
\label{}
\end{figure}
 
\noindent example, the minimum in 
$\rho_{ab}$ shifts from
$\approx 50\ K$ for the $x\ =\ 6.36$ sample to $\approx 175\ K$ for the $x=6.25$ sample.

\begin{figure}
\epsfxsize=\columnwidth
\epsfbox{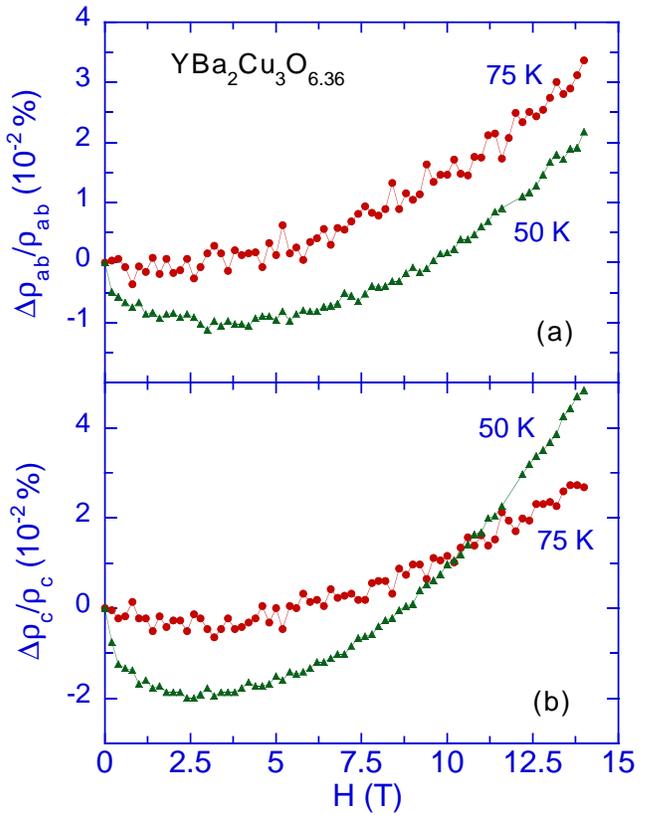}
\caption{Field $H$ dependence of magnetoresistivities for $H\parallel c$: (a)  $\Delta\rho_{ab}/\rho_{ab}$ and (b)
$\Delta\rho_{c}/\rho_{c}$ for $YBa_{2}Cu_{3}O_{6.36}$ at $50\ K$ and $75\ K$.}
\label{}
\end{figure}

 Measurements of the MR tensor on $YBa_2Cu_3O_{6.36}$ at temperatures higher than $75\ K $
showed that the field dependence of both in-plane and out-of-plane MR  is quadratic \cite {Houston}. 
However, at lower temperatures, the field dependence of the MR tensor becomes non trivial [see Figs. 3(a)
and 3(b)].  Namely, at  $50$ and $75\ K$, a negative  MR component, with
a large slope at relatively low fields, is superimposed on a positive quadratic field dependence. 
Notice that the minimum  in
$\Delta\rho_{c}/\rho_{c}$ is approximately twice the minimum in  
$\Delta\rho_{ab}/\rho_{ab}$. At higher magnetic fields, MR is positive and changes quadratically with $H$.

Figures 4(a) and 4(b) display the magnetic field
$H$ dependence of MR of the
$YBa_2Cu_3O_{6.25}$ single crystal.  The same phenomenon as in the previous sample, namely, sharply
developing negative MR in low fields  which turns into positive quadratic dependence in high fields, is even
more pronounced. 

The field dependence of MR illustrated in
Figs. 3 and 4 can be described for both samples as a sum of two contributions: a negative contribution
$Q_{i}(H)$ where $i=\{c,ab\}$, and a positive quadratic contribution; i.e.,
\be
\frac{\Delta \rho_{i}}{\rho_{i}}= Q_{i}(H)
+\gamma_{i} H^2;\;\;\;Q_{i}(H)<0; \gamma_{i} >0.
\ee

We examine in more detail  the field dependence of the MR components measured at
$75\ K$ on $YBa_2Cu_3O_{6.25}$ in Fig. 5 (a). First, we  note that the negative contribution to MR is much
larger for
the out-of-plane component than for the in-plane component; for example, the absolute value of
$\Delta\rho_{c}/\rho_{c}$ at its minimum is about
$7$ times the corresponding $\Delta\rho_{ab}/\rho_{ab}$. Second, 
both $\Delta\rho_{c}/\rho_{c}$ and $\Delta\rho_{ab}/\rho_{ab}$ measured in high magnetic fields are 
well fitted with  a parabolic dependence
$\gamma_{i} H^2 -\epsilon_{i}$, as shown by the solid curves, with
$\gamma_{c}=1.3\times 10^{-5}\;T^{-2}$, $\gamma_{ab}=6.25\times 10^{-6}\;T^{-2}$, $\epsilon_{c}
\approx 0.15\% $, and $\epsilon_{ab}
\approx 0.03\% $.  We discuss the origin of this quadratic $H$- dependence at the end of the next Section.

\begin{figure}
\epsfxsize=\columnwidth
\epsfbox{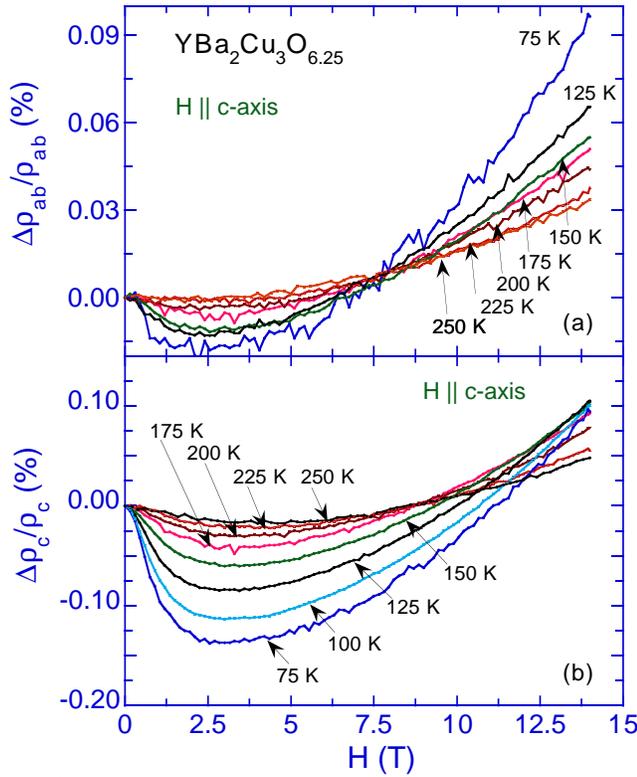}
\caption{Field $H$ dependence of  magnetoresistivities: (a) 
$\Delta\rho_{ab}/\rho_{ab}$ and (b)
$\Delta\rho_{c}/\rho_{c}$ for $YBa_{2}Cu_{3}O_{6.25}$.}
\label{}
\end{figure}

The field dependence of $Q_{i}(H)$ in Eq. (4) can be  determined by subtracting $\gamma_{i}
H^2$ from the total MR, and is illustrated in the semi-log plot of Fig. 5(b). The straight lines are fits to the data
and reveal the  logarithmic dependence of $Q_{i}(H)$ within a  certain range  of $H$. 
The functional dependence of $Q_{i}(H)$ can be summarized as follows:

\begin{eqnarray}
 Q_{i}  & \propto & \left\{ 
\begin{array}{lr} -\epsilon_{i}\  H^2/H_{0}^2  &  H< H_0 \\
  -\epsilon_{i}\ \frac{\ln (H/H_0)}{\ln(H_1/H_0)}\;\;\;\; & H_0 < H < H_1\\
-\epsilon_{i} & H> H_1
\end{array} 
          \right.  
\end{eqnarray}  
The most interesting part is the logarithmic dependence, covering almost a decade
of $H$. The value of the parameter $H_0$   is determined by the
intercept of the straight line according to Eq. (5). For $H> H_1\approx 4\ T$, $Q_{i}(H)$ saturates at the value
of $\epsilon_{i}$ obtained earlier by the parabolic fit in Fig. 5(a). The 
$\ln(H_1/H_0)$ in the denominator of Eq. (5) is required to match the interpolations of $Q_{i}(H)$ at 
$H_0< H< H_1$ and $H> H_1$. 

\begin{figure}
\epsfxsize=\columnwidth
\epsfbox{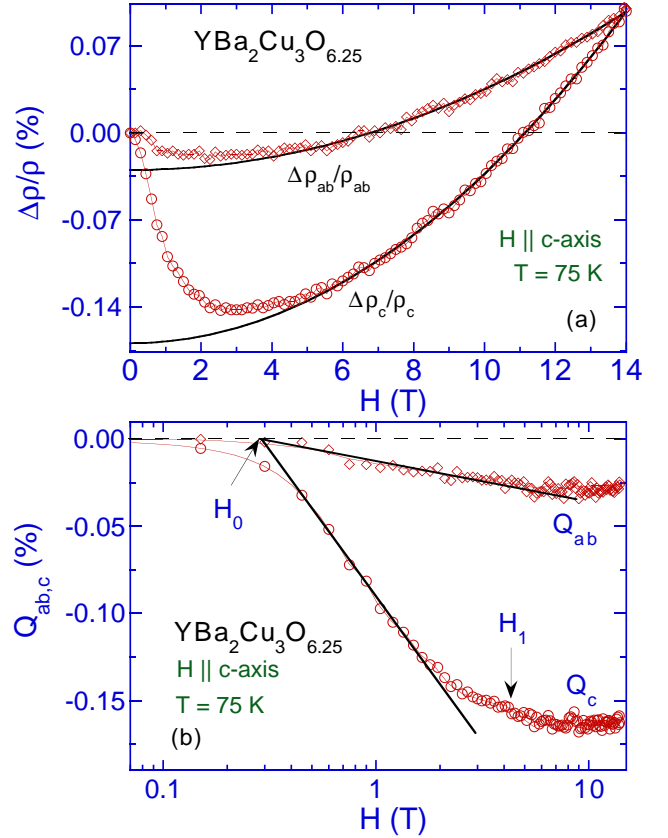}
\caption{(a)  Magnetoresistivities $\Delta\rho_{c}/\rho_{c}$ and $\Delta\rho_{ab}/\rho_{ab}$
for $YBa_{2}Cu_{3}O_{6.25}$ at $T=75\;K$ in  magnetic fields $H\| c$ . The solid lines
represent quadratic fits $\gamma_i H^2-\epsilon_i$ ($i=\{c,ab\}$).
(b) Orbital  component of magnetoresistivity $Q _{i}=\Delta\rho_{i}/{\rho_{i}}-\gamma_{i}H^2$ plotted vs
logarithm of field. The straight lines are guides to the eye corresponding to the logarithmic dependence given by
Eq. (5). The crossover fields $H_0$ and $H_1$ are indicated.}
\label{}
\end{figure}

The negative contribution to the in-plane MR is smaller
and, therefore, the data are noisier, but a similar procedure  of fitting the high field regime data with
a parabolic dependence yields  $Q_{ab}(H)$ with approximately the same H-dependence as 
$Q_c(H)$; i.e., $Q_c(H)/Q_{ab}(H)\approx const = 7$.

\section{ Discussion}

The most intriguing observation that emerges from the data presented above (Figs. 3 and 4) is the presence 
of a negative  MR component in low fields [given by $Q_{i}(H)$ in Eq. (4)], especially pronounced in
$\Delta\rho_{c}/\rho_{c}$, which exhibits at the lowest accessible temperatures a logarithmic $H$-dependence.
This negative MR is superimposed on a positive quadratic background [$\gamma_{i} H^2$ in Eq. (4)],
which is due to antiferromagnetic ordering \cite{Lavrov}. Below, we
show that the low field features of MR are direct consequences of interlayer incoherence.

Our approach is based on  the understanding that strongly underdoped layered crystals like
$YBa_2Cu_3O_{x}$  are characterized by incoherent out-of-plane transport. This was shown in 
optical measurements of the conductivity  on underdoped 
$YBa_2Cu_3O_{x}$\cite{Basov}.  In incoherent crystals, the value of the out-of-plane coherence length
$\ell_{\varphi ,c}$ is  T-independent, equal to its minimum possible value. Since the coherence length of
electron wave functions in a crystal cannot be smaller than the size of the atomic orbitals  
that overlap  over the
interbilayer  distance, the minimum possible value of
$\ell_{\varphi ,c}$ is the interbilayer distance; i.e., 
$\ell_{\varphi ,c}=\ell_0=11.7\AA$.  Thus,
a fundamental property of fully incoherent crystals is that $\ell_{\varphi ,c}$
does not  change with temperature or magnetic field. Therefore, the only  length scale which determines the
dissipation and can change with temperature or  applied magnetic field is the in-plane phase
coherence length $
\ell_{\varphi }$.
(Hereafter,  we
assume for brevity that the  planes are isotropic and omit the subscript $ab$.)
 Under these conditions, both conductivities depend only on the variable $ \ell_{\varphi }$
[$\sigma_{ab} (\ell_{\varphi }) $ and 
$\sigma_{c} (\ell_{\varphi })$], 
so that
their temperature and field dependences  come from that of $ \ell_{\varphi }$. 
(An instructive discussion regarding the dependence of conductivity on the spatial length scale of  
inelastic scattering processes is  presented in\cite{Abrikosov}).

The immediate consequence is a correlation between the magnetoconductivities and the
field variation of the in-plane coherence length at constant $T$:
\begin{equation}
\frac{\Delta \sigma_{ab} (H)}{\sigma_{ab}} = \kappa_{ab}
\frac{\Delta\ell_{\varphi}(H)}{\ell_{\varphi}};\;\;
\frac{\Delta \sigma_{c} (H)}{\sigma_{c}} = \kappa_{c}
\frac{\Delta\ell_{\varphi} (H)}{\ell_{\varphi}}.
\end{equation}
where $\kappa_{i}\equiv d \ln\sigma_{i}/d\ln\ell_{\varphi}$ ($i=\{ab,c\}$). 

The coefficients  $\kappa_{i}$ are dimensionless numbers of the order of unity. Their sign determines the
type of electrical conduction (metallic or nonmetallic); i.e., if $\kappa_{i}>0$, then
$\partial\sigma_{i}/\partial T <0$ and the conduction is metallic, while 
if $\kappa_{i}<0$, then $\partial \sigma_{i}/\partial T >0$ and the conduction is
nonmetallic\cite{note1}.
It is important to remember that in Eq. (6) and hereafter, $\Delta\sigma_{ab}/\sigma_{ab}$ and
$\Delta\sigma_{c}/\sigma_{c}$ are  the "orbital" contributions to magnetoconductivity
also denoted as $Q_i$ in Eq. (4).  

According to Eq. (6), the sign of the MR tensor is determined by the sign of $\kappa_{i}$ and by the effect of the
magnetic field on $\ell_{\varphi}$. It is known\cite{Altshuler}, and in the Appendix shown in more detail,
that the magnetic field decreases the phase coherence length due to its effect on self-intersecting
trajectories, so that $\Delta\ell_{\varphi}(H)<0$ in a weak field. The  
sign of $\kappa_{i}$ can be inferred from the $T$-dependence of the resistivity.  Specifically, 
for $YBa_2Cu_3O_{6.25}$, $\kappa_{c,ab}(T)<0$ for $T<175\ K$ since both $\rho_c$ and $\rho_{ab}$ are
nonmetallic [see Figs. 2(a) and 2(b)].  Then,  according to Eq. (6), both magnetoconductivity components are
positive (MR, correspondingly, are negative) for
$T<175\ K$, as indeed is the case [see Fig. 4]. By applying the same reasoning to the
$YBa_2Cu_3O_{6.36}$ sample, one concludes that $\Delta\rho_{c}/\rho_{c}$ should be negative, while 
$\Delta\rho_{ab}/\rho_{ab}$ should be practically zero at $75\;K$ and should become negative for
$T\leq 50\;K$,  where $\rho_{ab}$ turns nonmetallic as well. This is consistent with the low field data of Fig. 3. 

Also, according to Eq. (6), the field dependence of both MR is the same, determined by that of
$\Delta\ell_{\varphi}(H)$. Therefore,  their ratio should be a constant, given by the ratio of the
corresponding $\kappa_{i}$; i.e.,
\be
\frac{\Delta (\ln\sigma_{c})}{\Delta (\ln\sigma_{ab})}=\frac{\kappa_c}{\kappa_{ab}}.
\ee
As discussed in
Refs.\cite{e-print,Almasan}, in incoherent crystals 
$\sigma_c(\ell_{\varphi })\propto \sigma_{ab}(\ell_{\varphi })/\ell_{\varphi }^2$.
As a result, $\kappa_c =\kappa_{ab} -2$. Therefore, when  $\kappa_{ab}$ is negative
(as is the case with our crystals at low temperatures), the absolute value of   $\Delta
(\ln\sigma_{c})$ is greater than that of $\Delta (\ln\sigma_{ab})$, namely:
\be
\frac{\Delta (\ln\sigma_{c})}{\Delta (\ln\sigma_{ab})}=\frac{|\kappa_{ab}|+2}{|\kappa_{ab}|}.
\ee
In other words, when the T-dependence of both conductivities is nonmetallic, the out-of-plane 
conductivity depends stronger on the phase coherence length, and the respective MR is greater by a
constant factor. 

This conclusion is, indeed, supported by the experimental data.  The ratio of the magnetoresistivities is a
constant factor which can be estimated, for example, from Fig. 5(b). Namely, for
$YBa_2Cu_3O_{6.25}$ at
$T=75\;K$, $Q_{c}$
is about seven times greater than $Q_{ab}$. 
From Eq. (8), we estimate  $\kappa_{ab}\approx   -0.33$. 
 
We turn now our attention to the specific field dependence of the magnetoresistivities. The
logarithmic field dependence of Q at low temperatures and in relatively small magnetic fields indicates
2D quantum interference \cite{Altshuler,Abrikosov}. This effect is due to the 
presence of self-intersecting trajectories along which electrons can traverse the loop in opposite
directions\cite{Altshuler,Abrikosov}. A small applied magnetic field gradually destroys the
quantum interference, which induces a
variation in the phase coherence length and, therefore, a variation in the conductivity.

 The self-intersecting trajectories constitute a small fraction of the total number of phase coherent
trajectories of a given length. The great majority  of the phase coherent trajectories
(trajectories over which the phase changes predictably and,
 therefore,
no irreversible processes are involved) are not self-intersecting. 

As Fig. 2 shows, both resistivities change strongly with temperature and, therefore, their
magnitude and T-dependence are determined by the contribution of the majority, non-intersecting
trajectories. 
The small, temperature dependent corrections to 
conductivities (in zero field)  due to self-intersecting trajectories, known as
corrections due to weak localization, can only be observed when the main contribution to conductivity which is
due to non-intersecting  paths saturates (regime of residual resistivity). Our samples are not in this regime.
Therefore, the contribution of the trajectories with
loops to the T-dependence of both conductivities is negligible in our samples  and  $\sigma_{ab}(T)$
and $\sigma_{c}(T)$ do not reflect the characteristic $\ln T$  (in 2D) or
$T^{1/2}-T^{1/3}$ (in 3D) dependence due to quantum interference, which is usually the subject of
discussion  in the literature. 

On the other hand,  
the magnetoconductivity
produced by a {\it weak}  magnetic field is due to   self-intersecting trajectories because all the other
(conventional) contributions to magnetoconductivity are still negligible.   Therefore, the magnetoconductivity
caused by the effect of the field on the self-intersecting trajectories can be  experimentally identified.  

In order to determine the field dependence of
both magnetoconductivities, we need to evaluate the magnetic field variation  of the
coherence length [$\Delta\ell_{\varphi}(H)/\ell_{\varphi}$]. 
A semi-quantitative derivation of
this dependence is given in the Appendix, with the following results:

\begin{eqnarray}
 \frac{\Delta\ell_{\varphi}}{\ell_{\varphi}}  & \propto & \left\{ 
\begin{array}{lr} -\eta\;H^2/H_0^2  &  H< H_0 \\
  -\eta\;\frac{\ln (H/H_0)}{\ln(H_1/H_0)}\;\;\;\; &  H_0 < H < H_1\\
-\eta & H> H_1 
\end{array} 
          \right.  
\end{eqnarray}  
Here $\eta $ is a small number determined by the relative weight of the number of the
self-intersecting trajectories with respect to that of the majority nonintersecting trajectories.  The
lower crossover field $H_0$ is determined by the condition that the field flux through the largest
loops along the  phase coherent trajectories is approximately equal to the flux quantum
$\phi_0=2\times 10^{-7} Oe\;cm^2$; namely $H_0\sim \phi_0/ \ell_{\varphi }^2$. The upper crossover
field $H_1$ is determined by the similar condition that the field flux is of the order of
$\phi_0$ through the smallest possible loops  (of the order of the elastic mean free path $l_{el}$); 
i.e., $H_1\sim \phi_0/ l_{el}^2$. For $H>H_1$, no trajectory with loops contributes 
to the value of the  phase coherence length. Hence, $\Delta\ell_{\varphi}$ saturates.

From Eqs. (6) and (9) follows Eq. (5)  with $\epsilon_{i}= -\eta\ \kappa_{i}$. 
The nontrivial logarithmic field dependence appears only when there is a significant field range
between $H_0$ and $H_1$. This requires  the phase coherence length  to be substantially greater
than the elastic mean free path, which means a   well developed diffusive regime.
Therefore, the magnetoeffects due to self-intersecting trajectories are more pronounced 
in stronger underdoped samples. 
For $YBa_2Cu_3O_{6.25}$ at $T=75\;K$ [see Fig. 5(b)], we estimate
that $\epsilon_{c}\approx 0.15\%$, $H_0\sim 0.3 \;T$ which corresponds to $\ell_{\varphi
}\sim 10^3\;\AA$, and $H_1\sim 4\;T$ ($\ell_{el}\sim 270\AA$).   For $H>4\;T$, the
contribution of quantum interference to magnetoconductivities saturates and   
other types of contributions to MR become dominant, as discussed below. 

Experimentally, one can relatively easy distinguish between the cases of two-dimensional and
three-dimensional phase coherent trajectories. First, the magnetoeffect of self-intersecting trajectories is
substantially greater in 2D, because  of the greater probability of a 2D trajectory to form a large loop
($\eta\sim \lambda/\ell_{el}$ in 2D and $\eta\sim\lambda^2/\ell_{el}^2$ in 3D, see the Appendix).
Therefore, in this case, the effect becomes pronounced and observable at  relatively high
temperatures.   Second, the logarithmic $H$ dependence can be distinguished from $H^{1/2}$ rather 
clearly even for only one decade in $H$. 
 These features observed in MR would point to the 2D nature of the phase coherent trajectories. 
The two manifested together, as is the case with our MR data of $YBa_2Cu_3O_{x}$ shown above,
make this conclusion very compelling. It should be noted that the "three dimensional" $H^{1/2}$
dependence of MR in manganites was reported recently by Qing'An Li, Gray, and Mitchell\cite{Gray}.

Finally, we discuss briefly the origin of the  positive contribution that  changes as
$\gamma_i H^2$ and becomes dominant at applied magnetic fields larger than $4\;T$ (see Figs.
3 and 4). A similar type of MR  was reported by other groups as well \cite {Lavrov}, and it was associated with
the antiferromagnetic fluctuations and antiferromagnetic ordering. Indeed,
$YBa_2Cu_3O_{6.25}$  is antiferromagnetic for all temperatures below  room temperature,
while  $YBa_2Cu_3O_{6.36}$ is antiferromagnetic
for $T\leq 40\;K$. In high
magnetic fields, the contribution of the AF correlations to MR  dominates the smaller contribution of the
self-intersecting trajectories. At sufficiently low temperatures, these two
contributions can be clearly separated as illustrated in Figs. 5(a) and 5(b).

\section{ summary}

We present magnetoresistivity data for two strongly underdoped single crystals of $YBa_2Cu_3O_{x}$
with $x=6.36$ and $6.25$. Both in-plane  $\Delta\rho_{ab}$ and  out-of-plane $\Delta\rho_{c}$
magnetoresistivities MR were measured simultaneously on the same single crystal using a
multiterminal method. This permits a direct comparison between their temperature and field
dependences.  

The most interesting observation is a negative MR  in low applied
magnetic fields.
This low field contribution is characterized by two important features: First, the effect is strong and, therefore,
is pronounced even at relatively high temperatures ($75\;K$ and higher). Second,
the field dependence is consistent with $\ln (H)$, rather than $H^{1/2}$. These  features point    towards  the two dimensional
nature of the phase coherence  in these crystals; namely, the phase-coherent volume contains only one or two neighboring
bilayers, while the coherence length along the planes is orders of magnitude greater than the size of the unit cell.

We attribute the second contribution to MR,  which changes quadratically with field up to  $H=14\;T$,
to the effect of $H$ on the AF correlations. This positive contribution dominates at fields
above $4\;T$ and induces a change in the sign of the total MR for both components of the magnetoresistivity.

\section{ appendix}

To determine the effect of the magnetic field on self-intersecting phase coherent trajectories, and the
resultant effect on the phase coherence length, we use the qualitative approach 
described in\cite{Altshuler,Abrikosov}.
The coherent trajectories consist of two groups. The great majority of them are not self-intersecting
and they determine the value and the  temperature dependence  of  the phase coherence length and, 
consequently,
of the conductivities. A much smaller fraction of the coherent trajectories have loops
and exhibit the interference effect. 
 
We begin our estimate by defining the phase coherence length in zero magnetic field
$\ell_{\varphi}(0)$ as the mean square average distance in the x-direction that an electron travels while
retaining its  phase memory (i.e., without encountering dephasing inelastic collisions):  
\begin{equation}
\ell_{\varphi}^2(0)=\int_{0}^{\infty}x^2[(1-\eta )P_0(x)+\eta P_{si}(x)]dx.
\end{equation}
Here $P_0(x)$ [$P_{si}(x)$] is the probability that the phase coherence is retained between points $A$
and $B$  separated by a  distance $x$, when  all trajectories without [with] self-intersections are
counted. Both probabilities are normalized to
unity. The small number
$\eta \ll 1$ reflects the relative weight of self-intersecting trajectories. We also introduce
$p(x,{\cal A})$ as the probability density that a self-intersecting, coherent trajectory  (in zero field) has a
loop of area ${\cal A}$, subject to  the normalization condition:
\begin{equation}
\int_{\pi\ell_{el}^2}^{\pi x^2}p(x,{\cal A})d{\cal A}=P_{si}(x).
\end{equation}
Here, the smallest possible area of the loop is determined by the elastic mean free path $\ell_{el}$,
and the largest possible loop is determined by the distance  $x$ between $A$ and $B$.

Along self-intersecting trajectories, an electron can traverse the loop
in opposite directions with no phase difference between the respective amplitudes. 
An applied  magnetic field, however,  introduces a phase difference 
$\varphi_H=2\pi\phi/\phi_0$ between the two alternative routes along the same trajectory.
Here   $\phi ={\cal A}H $ is the  magnetic field flux through the loop, ${\cal A}$ is the area of the loop
normal to the field, and 
$\phi_0=2\times 10^{-7} Oe\;cm^{2}$ is the flux quantum. As a result,  when an electron arrives at point $B$,
following a self-intersecting route, its phase is unpredictable, and varies by $\varphi_H \ge\pi$   if the
corresponding trajectory has a loop with area ${\cal A}\ge\phi_0/2H$. Therefore, such a trajectory is
no longer phase coherent and  does not contribute to the average given by Eq. (10). This is the
mechanism by which a weak field  changes the coherence length and, respectively, the conductivities. 

We give a rough estimate of $\Delta\ell_{\varphi}(H)$ by considering that 
all self-intersecting trajectories with loop areas ${\cal A}\ge \phi_0/2H$ do not
contribute to the average in Eq. (10), while those with ${\cal A}<\phi_0/2H$ still do.
 The phase coherence length in a magnetic field is then given by:
\begin{equation}
\ell_{\varphi}^2(H)=\int_{0}^{\infty}x^2dx\left [(1-\eta )P_0(x)+\eta
\int_{\pi\ell_{el}^2}^{\pi\ell_{H}^2}p(x,{\cal A})d{\cal A}\right ],
\end{equation}
where the magnetic length $\ell_{H}=(\hbar
c/2eH)^{1/2}$ is determined by the  condition $\pi\ell_{H}^2=\phi_0/2H$.

Taking into account Eqs. (10) and (11) and the small value of $\eta $, we obtain from Eq. (12):
\begin{equation}
\frac{2\Delta\ell_{\varphi}(H)}{\ell_{\varphi}}\approx
-\eta\frac{\int_{\ell_{H}}^{\infty}x^2dx\int_{\pi\ell_{H}^2}^{\pi x^2}p(x,{\cal A})d{\cal A}}
{\int_{0}^{\infty}x^2P_0(x)dx}.
\end{equation}
Thus, the {\it decrease } of the phase coherence length $\Delta\ell_{\varphi}(H)$  is proportional to
the weight of excluded  self-intersecting trajectories, namely, those with loop areas 
${\cal A }\ge\pi\ell_{H}^2$.  The lower
limit of integration  in $x$ is $\ell_{H}$ because the trajectories with a coherent distance
$x\le\ell_{H}$ cannot have loops with area ${\cal A}\ge \pi\ell_{H}^2$ and, therefore,
are not part of the excluded  self-intersecting trajectories. 

Next we find the expression for $p(x,{\cal A})$ using the
diffusive approximation\cite{Abrikosov}.  Since phase-coherent trajectories are two-dimensional (we
discuss here incoherent crystals), the probability to  find an electron at a point
$\vec r$ from its starting origin at $t=0$,  is $W(\vec r )d^2r\propto \exp (-\vec
r^2/4Dt)d^2r/4Dt$, where $D$ is the difussion coefficient. The trajectory intersects itself between the times $t$
and
$t+dt$ if it enters the area
$\lambda vdt$ around the origin. Here $\lambda$ is the de Broigle wavelength (the trajectory is
viewed  as a tape of width $\lambda$, rather than a line). 
The probability of return, and, consequently, the probability to form a loop is given by
$p(t)dt\propto W(0)\lambda vdt\propto \lambda vdt/Dt$. 
Since   the area of the loop is proportional to the average 
distance  from the origin,   ${\cal A}\propto \langle\vec r^2\rangle\propto Dt$, and 
$d{\cal A}\propto Ddt$, the probability of a loop with area between ${\cal A}$ and ${\cal A}+d{\cal A}$
is given by $(\lambda v/D)d{\cal A}/{\cal A}$. 
Thus,  the probability that a 2D trajectory has a loop of area ${\cal A}$ scales as $1/ {\cal A}$.

The prefactor $ \lambda v/D$ determines the overall probability of self-intersecting trajectories and
gives the number $\eta$, which we introduced "by hand" earlier: 
$\eta\sim \lambda v/D \sim \lambda/\ell_{el} $, because $D\sim v \ell_{el}$.
Assuming $\lambda \sim 2-3\;\AA$ and $\ell_{el}\sim 200-300\;\AA$, which is a reasonable estimate
given the density of charge carriers, we obtain $\eta\sim 0.01$. 
It means that the maximum effect of removing self-intersecting trajectories from the average in Eq. (13) reduces
$\ell_{\varphi}$ by about one percent. Other numerical factors also  absorbed in $\eta$ can reduce it
further. This correlates with the maximum value of  the negative MR in Fig. 5(b) of the order of $0.1\%$.

Since the diffusive motion has no memory, having a loop does not affect the statistical properties of the rest of
the trajectory. Therefore, the probability density
$p(x,{\cal A})d\cal A$ can be roughly estimated as:
\be
p(x,{\cal A})d{\cal A}\approx \frac{P_{si}(x)}{\ln (x^2/\ell_{el}^2)}\frac{d{\cal A}}{{\cal A}},
\ee
where the logarithm in the denominator is the normalization factor due to Eq. (11). Now the integral
in (13) becomes:
\be
\int_{\ell_{H}}^{\infty}x^2dx\int_{\pi\ell_{H}^2}^{\pi x^2}p(x,{\cal A})d{\cal A}
= \int_{\ell_{H}}^{\infty}x^2P_{si}(x)\frac{\ln (x^2/\ell_{H}^2)}{\ln (x^2/\ell_{el}^2)}dx.
\ee

The characteristic logarithmic field dependence of this integral appears when
$\ell_{H}^2$ decreases below the value of $\ell_{\varphi }^2$. 
This is equivalent to the condition    that the
field flux through the largest loops along the  phase coherent trajectories is approximately equal 
to $\phi_0/2$; hence, the characteristic field $H_0= \phi_0/ 2\pi\ell_{\varphi }^2$. 
As $\ell_{\varphi }$ increases with decreasing
temperature, the  onset field $H_0$ can become very small and the MR due to the interference effect 
manifests itself when all the other sources of MR are negligible. The upper crossover field is defined by the
condition that the field flux is $\phi_0/ 2$ through the smallest loops with the area of the order of
$\pi\ell_{el}^2$; hence $H_1=\phi_0/ 2\pi\ell_{el }^2$. The upper field does not change with temperature,
while the lower field $H_0$ decreases with decreasing $T$. Therefore, this effect is observable at sufficiently low
temperatures when $H_0< H< H_1$ or, equivalently, 
$\ell_{\varphi }^2>\ell_H^2> \ell_{el }^2$.

In this regime, the integral (15) can be easily estimated because  
the  logarithms are slowly changing functions and they can be taken at the value $x\approx\ell_{\varphi}$,
which corresponds to the maximum in $x^2P_{si}(x)$. Other contributions to this integral are negligible
in comparison with the logarithmic increase $\sim \ln (\ell_{\varphi }^2/\ell_H^2)=\ln (H/H_0)$.
Correspondingly, the logarithm in the denominator of Eq. (15) becomes
$ \ln (\ell_{\varphi }^2/\ell_{el}^2)=\ln (H_1/H_0)$. In this regime, the
denominator in Eq. (13), $\int_{0}^{\infty}x^2P_0(x)dx\sim
\int_{0}^{\infty}x^2P_{si}(x)dx\approx \ell_{\varphi }^2$,  and we get Eq. (9), with all numerical factors
absorbed in $\eta$. Hence, according to Eq. (6), both conductivities have the logarithmic field dependence.
For $H>H_1$, the integral (15), hence, Eq. (13) saturates since no self-intersecting
trajectories  contribute to the average (12). 

It is important to underline   that the transport in these crystals is  obviously
three-dimensional, but the phase coherence is two-dimensional, 
meaning that the distribution of the  area of the loops along phase coherent trajectories
scales as  $1/{\cal A}$, leading to the logarithmic field dependence of $\Delta\ell_{\varphi}(H)$ and both
magnetoconductivities. This result remains valid even when the coherence is established between
two neighboring bilayers. As long as $\ell_{\varphi,c }\sim\ell_0\ll \ell_{\varphi,ab }$, the distribution of loop
areas is close to $1/{\cal A}$.

Truly 3D phase coherent trajectories requires $\ell_{\varphi,c }\gg \ell_0$. Then, a similar
analysis\cite{Abrikosov} shows that the distribution of the loop areas is $d{\cal A}/{\cal A}^{3/2}$ 
and $\eta\sim\lambda^2/\ell_{el}^2$. Then, in the regime $H_0\ll H\ll H_1$, Eq. ($A$) gives
\be
\frac{\Delta\ell_{\varphi}(H)}{\ell_{\varphi}}\sim
-\frac{\lambda^2}{\ell_{el}^2}\frac{\ell_{el}}{\ell_{H}}\sim -\frac{\lambda^2}{\ell_{el}^2}
\left (\frac{H}{H_1}\right )^{1/2}
\ee

{\it This research was supported by the National Science Foundation under
Grant No.
DMR-9801990 at KSU and the US Department of Energy under  Contract No.
W-31-109-ENG-38 at ANL}.


\begin{references}
\bibitem{note1} Indeed, since in-plane
$\sigma_{ab}$ and out-of-plane $\sigma_{c}$ conductivities are solely functions of
$\ell_{\varphi}$, the following equation holds:
\be
 \frac{\partial \ln\sigma_{c,ab}}{\partial T}=\kappa_{c,ab}\frac{\partial\ln\ell_{\varphi}}{\partial T}.
\ee
Since the coherence length always
decreases with increasing temperature,  i. e. $\partial\ell_{\varphi}/\partial T<0$, the sign of $\kappa_{i}$
determines the sign of $\partial\sigma_{i}/\partial T$, hence the type of conduction.
\end{references}
\end{document}